# Transcutaneous Spinal Cord Stimulation Disrupts Conscious Ankle Proprioception and Produces a More Constrained Locomotor Pattern in Unimpaired Adults

Christopher A. Johnson, Andria J. Farrens, Parastoo Ali Pour, Arjan Gillan, Hui Zhong, David J. Reinkensmeyer, Alexandra S. Voloshina

***Abstract*— Transcutaneous spinal cord stimulation (tSCS) modulates spinal sensorimotor circuits primarily through activation of afferent networks. While prior work has emphasized locomotor performance and spinal excitability, how tSCS affects conscious proprioceptive perception and the extent to which such effects parallel changes in locomotor control remain unclear. We investigated the acute and training-related effects of tSCS on ankle proprioception and gait in unimpaired adults (n = 14), with an independent control group (n = 14) completing identical proprioceptive training without stimulation. Proprioception was quantified using a bilateral robotic assessment of dynamic ankle localization ability (Crisscross), gross motor output using maximum dorsiflexion strength, and gait during normal and tandem treadmill walking using spatiotemporal, trunk-sway, and mediolateral center-of-mass (CoM) excursion measures. Acute tSCS increased ankle proprioceptive error ($p < 0.001$) while dorsiflexion strength was unchanged ($p = 0.30$). Gait shifted toward a modestly more constrained locomotor pattern, characterized by reduced step width and ML CoM excursion ($p < 0.05$). With continued training under stimulation, proprioceptive error decreased and, unlike the control group, the tSCS group showed progressive improvement that persisted after stimulation ended. Sagittal-plane gait measures recovered toward or beyond baseline, whereas mediolateral measures remained constrained, revealing a direction-dependent reorganization of locomotor control. Together, these findings show that tSCS influences multiple aspects of the sensorimotor control loop, disrupting conscious proprioception while reshaping locomotor behavior, and that the nervous system can adapt to altered afferent input through training.**

***Index Terms*— Gait, Neuromodulation, Proprioception**

## I. Introduction

Proprioception, the ability to sense limb position and movement, is key for motor learning, coordination, and recovery [1], [2]. Proprioceptive information from muscles, tendons, joints, and skin provides the nervous system with continuous feedback regarding body position and movement, enabling accurate state estimation and ongoing adjustments to motor output [1], [2]. During locomotion, proprioceptive feedback contributes to the regulation of limb trajectories, balance control, and adaptive responses to changing environmental demands [3]. Impairments in proprioception have been associated with deficits in movement coordination, reduced mobility, and diminished rehabilitation outcomes across multiple neurological populations [3], [4]. While many of these functions rely on automatic sensorimotor processing, the assessment of proprioception in such studies is assessed through tests of conscious perception of limb position and movement, evidencing the predictive power of such tests. Conscious proprioception sense relies predominantly on the dorsal column medial lemniscus pathway, which relays proprioceptive information to the primary somatosensory cortex [3]. Such conscious proprioceptive ability is modifiable with training [4], [5]. However, the extent to which changes in conscious proprioceptive ability reflect, contribute to, or adapt alongside motor behavior remains poorly understood, particularly in the context of interventions aimed at modulating sensorimotor function.

Locomotion is often regarded as an automated motor behavior, but effective gait control emerges from dynamic interactions between spinal and supraspinal sensorimotor processes [6]. At the spinal level, afferent input contributes to reflex modulation, interneuronal processing, and the generation of coordinated locomotor output [7]. In parallel, supraspinal systems integrate sensory information to support conscious proprioceptive perception, state estimation, motor planning, and adaptive control [8]. These processes continuously interact during walking, allowing the nervous system to adjust motor behavior in response to changing sensory conditions. Because locomotor control depends on the ongoing integration of spinal and supraspinal afferent input, interventions capable of modulating such input could provide a valuable framework for investigating how gait control depends on sensory processing.

Transcutaneous spinal cord stimulation (tSCS) has emerged as a powerful neuromodulatory technique capable of altering spinal network excitability and enhancing voluntary motor output during standing and walking [9], [10]. Numerous studies have demonstrated that tSCS can facilitate muscle activation, improve stepping performance, and restore locomotor patterns in individuals with neurological injury, including stroke [11], [12], [13] and spinal cord injury [14], [15], [16]. These motor benefits are generally attributed to the recruitment of large-diameter afferents and modulation of spinal interneuronal circuits [9]. Because locomotor coordination depends heavily on the integration of afferent feedback, tSCS has primarily been interpreted as enhancing motor control through its effects on afferent processing and spinal circuitry [9], [10].

Despite the known effects of tSCS on motor performance, considerably less is known about how spinal neuromodulation influences conscious proprioceptive processing and its relationship to locomotor behavior. Although numerous studies have demonstrated improvements in gait [11], [17], [18], [19], [20] and balance control [21], [22] during spinal stimulation in both neurologically impaired and unimpaired populations, these outcomes do not reveal how stimulation alters conscious perception of limb position and movement. Because tSCS acts primarily by recruiting proprioceptive afferents, it is plausible that it also influences the sensory processes those afferents support, yet this possibility remains largely untested. To our knowledge, only two studies have directly quantified conscious proprioception during spinal stimulation, reporting opposing effects. One study found that stimulation preserved static position-matching accuracy while improving dynamic movement-matching performance of the upper limb [23]. In contrast, another study using epidural stimulation found that stimulation blocked a substantial portion of proprioceptive afferent input and reduced or abolished the conscious perception of leg position [24]. Because these findings diverge in both direction and stimulation modality, the effect of tSCS on conscious ankle proprioception cannot be inferred from prior work and remains an open question.

Beyond its immediate effects on proprioception and locomotor behavior, tSCS may also influence how sensorimotor function changes with practice. Specifically, as mentioned above, proprioceptive perception is not static and can improve through feedback-driven perceptual learning and sensorimotor recalibration [4], [5], [25]. Because sensory feedback provides critical information for state estimation and error correction, interventions that alter afferent processing may influence not only immediate proprioceptive performance but also the rate and extent of proprioceptive learning. By modulating the excitability of afferent pathways involved in sensorimotor processing, tSCS may facilitate error detection, sensory integration, and feedback-driven learning, potentially enhancing the magnitude of improvements in proprioceptive accuracy. Conversely, evidence from epidural spinal stimulation suggests that artificially increasing afferent activity may disrupt sensory precision [24], yet it is unclear if the motor system can learn to compensate for such disruptions. Therefore, the extent to which tSCS influences and/or depends on proprioceptive learning remains unclear.

The purpose of this study was to determine how modulating afferent input via tSCS affects multiple aspects of the sensorimotor control loop, including conscious ankle proprioception, locomotor behavior, and sensorimotor adaptation following proprioceptive training. In this initial study, we focused on unimpaired adults, to establish a normative baseline for the putative effects. We used a robotic ankle proprioception assessment to quantify conscious proprioceptive ability and gait analyses during normal and tandem walking to characterize locomotor behavior. Using these measures, we evaluated the acute effects of tSCS on proprioceptive perception and walking, as well as changes in proprioceptive learning following feedback-based training during stimulation. Based on the prevailing account that tSCS enhances sensorimotor function through recruitment of large-diameter afferents, we hypothesized that: (1) tSCS would disrupt conscious ankle proprioceptive accuracy by superimposing stimulation-evoked afferent activity onto naturally occurring proprioceptive signals; (2) tSCS would alter locomotor behavior, particularly during tandem walking, by promoting a more constrained control strategy in response to changes in afferent processing involved in balance regulation and postural control; and (3) proprioceptive training with tSCS will accelerate proprioceptive learning, relative to training without stimulation, by facilitating sensory integration and feedback-driven perceptual learning.

## II. Methods

We recruited 14 unimpaired participants (mean ± SD age: 26.1 ± 8.7 years; M/F: 9/5). Participants completed proprioception, walking, and motor function assessments with and without tSCS, as well as proprioceptive training with tSCS. We additionally analyzed data from a separate cohort of 14 unimpaired adults who completed the same proprioceptive assessment and training protocol without stimulation (control group; age: 25.5 ± 5.8 years; M/F: 5/9). Both cohorts were tested on the same robotic device, with identical task parameters and training procedures, differing only in the application of tSCS. The groups did not differ significantly in age or baseline proprioceptive accuracy ($p > 0.2$). Exclusion criteria included a history of neurological injury, musculoskeletal damage, or any injury that impaired ankle movement or sensation. All participants provided written informed consent before participation. All studies were approved by the University of California, Irvine Institutional Review Board (IRB #5889).

### *Robotic Assessments: Proprioception and Gross Motor Function*

We assessed ankle proprioception and dorsiflexion maximum voluntary contraction (MVC) using a previously developed robotic platform, Ankle Measuring Proprioceptive Device (AMPD) (Fig. 1A) [26], [27], [28]. AMPD is a bilateral robotic platform with two selectable impedance states, fully backdriveable and rigidly, motor-driven, in which case it can independently move each foot about the ankle joint to evaluate proprioceptive ability and ankle motor function. Participants sat in an upright position with their hips and knees flexed to 90°s, and their feet hip-width apart. Prior to securing the feet to the footplates, we added wood shims under the feet as needed to align the lateral malleolus with the rotational axis of AMPD. Participants then completed two robotic assessments: a proprioceptive assessment (Crisscross) and dorsiflexion maximum voluntary contraction (MVC). Standardized instructions and a demonstration were provided before each assessment.

The Crisscross assessment quantifies dynamic ankle localization ability by asking participants to indicate the moment when they perceive their left and right ankles to be at the same angular position. During the assessment, AMPD is placed in rigid mode and rotates the participant's passive ankles in opposite directions, producing a single crossing point. The assessed range of motion was fixed at 50° for all participants, spanning 15° of dorsiflexion to 35° of plantarflexion, and the crossing point was fixed within the plantarflexed region (~7° plantarflexion) across all trials (Fig. 1B). Participants were tested at three crossing speeds (16, 26, and 36 °/s), defined as the sum of the angular velocities of both ankles. Participants were exposed to six crossing trials at each speed (18 trials total), with no visual feedback, and were instructed to press a handheld button when they perceived their ankles to be aligned. To reduce predictability, movement speeds were pseudo-randomized across trials.

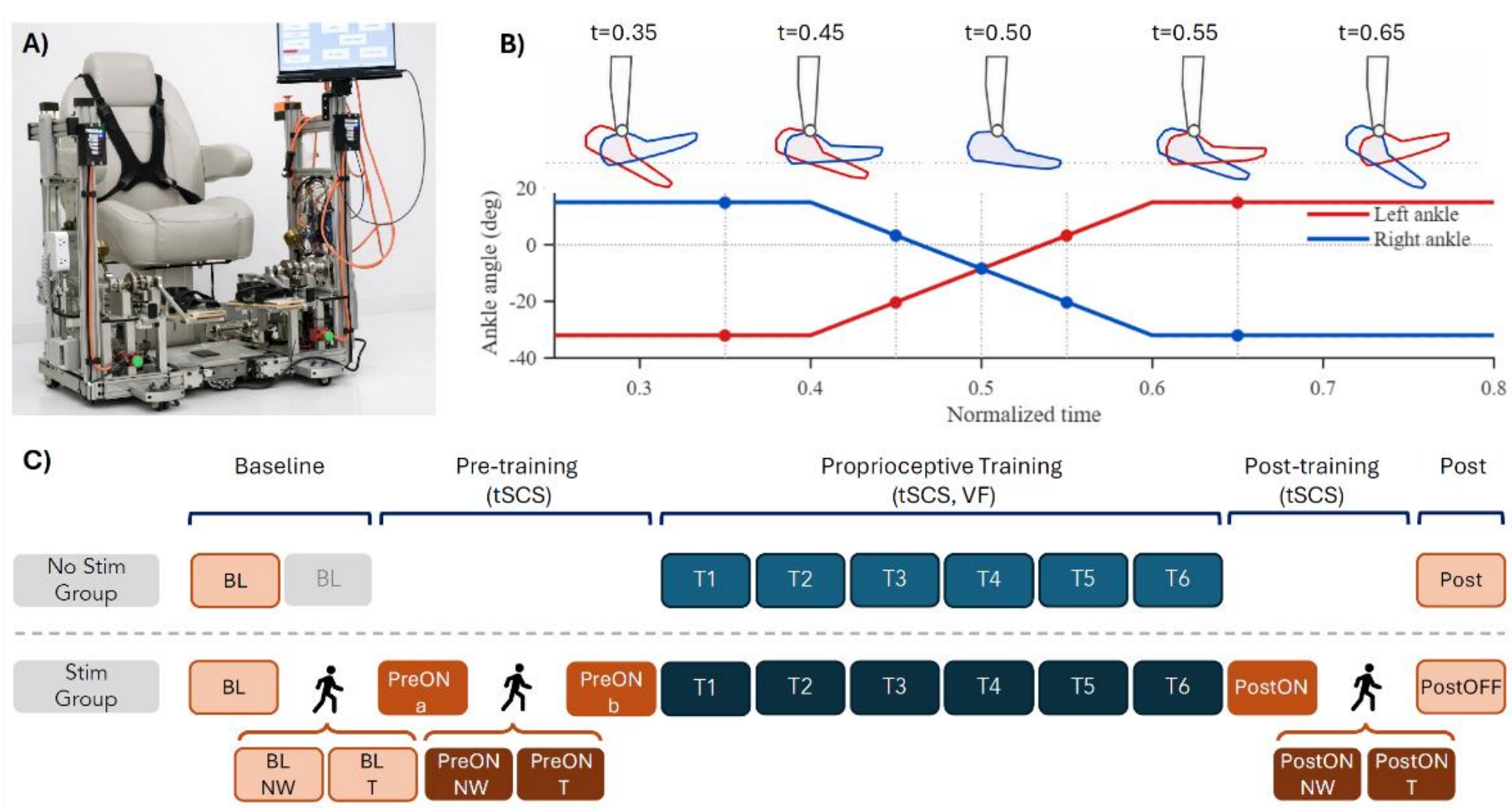

**Figure 1. Experimental device, proprioceptive assessment task, and study protocol. (A)** The robotic ankle device (AMPD) used to assess ankle and motor function. **(B)** Representative single crossing trial from the Crisscross assessment. Top: ankle configurations at successive normalized time points spanning the crossing. Bottom: representative left (red) and right (blue) ankle angle trajectories versus normalized time, with the crossing occurring in the plantarflexed region at approximately ~7° of plantarflexion. **(C)** Study protocol for the Control (top) and experimental group (tSCS; bottom). Both groups completed a Baseline block, six training blocks (T1–T6) with visual feedback [VF], and a Post assessment. The Stim group additionally completed pre-training (PreON [a, b]) and post-training (PostON) assessments during stimulation, with tSCS delivered throughout the Pre-training, Training, and Post-training phases. Walking icons denote treadmill gait assessments; BL NW / BL T, PreON NW / PreON T, and PostON NW / PostON T denote normal-walking (NW) and tandem-walking (T) assessments at each time point; PostOFF, post-training assessment with stimulation off

For the dorsiflexion MVC task, AMPD operated again in rigid mode, but with the ankles locked at 90° in the sagittal plane. Participants were asked to gradually increase dorsiflexion effort of their dominant ankle until they reached maximum effort and held the contraction for three seconds. Each participant completed two MVC trials separated by a 10 s rest, and the mean peak torque across both trials was used for analysis. We determined leg dominance by asking participants which leg they would use to kick a ball [29].

### Gait Trials: Normal and Tandem Walking

Participants completed gait trials on a dual-belt instrumented treadmill (Bertec Corporation, Columbus, OH, USA) at a fixed walking speed of 1.0 m/s. We evaluated two walking conditions: normal walking and tandem walking. During normal walking, participants were instructed to walk naturally at the prescribed speed. During tandem walking, participants were asked to place one foot in front of the other while walking on top of a strip of yellow tape positioned on the left treadmill belt. The tape served as a visual guide to standardize foot placement and maintain a narrow base of support throughout the task. Tandem walking was included because it increases reliance on sensorimotor integration and postural control by reducing the base of support, making it well suited for evaluating locomotor adaptations to altered afferent input [30]. During a familiarization period, participants were permitted to look at their feet while practicing the task. Once data collection began, we instructed participants to look forward and avoid looking at their feet unless necessary.

Each walking condition lasted three minutes. We recorded ground reaction forces from the instrumented treadmill and kinematic data using an 8-camera Vicon motion capture system (100Hz; Vicon Motion Systems Ltd., Oxford, UK). Reflective markers were placed bilaterally on lower-extremity anatomical landmarks according to VICON's Plug-in Gait lower body model and over the spinous process of C7 to quantify lower-limb kinematics and trunk motion.

### Transcutaneous Spinal Cord Stimulation (tSCS)

We delivered continuous stimulation using the BioStim 5 system (Cosyma Inc., Moscow, Russia) with a biphasic waveform (10-kHz carrier frequency, 30-Hz burst frequency). The stimulation parameters selected for this study have been shown to effectively engage spinal circuits involved in locomotor control in people with spinal cord injury [31], [32], [33]. Three 1.25-in circular cathodes (LotFancy, Inc., San Francisco, CA, USA) were positioned over the C5–C6, T11–T12, and L1–L2 interspinous spaces, while two 2 × 4-in rectangular anodes (Reserv, Inc., Hudson, OH) were placed over the bilateral iliac crests. The thoracic and lumbar stimulation sites shared a common reference electrode on the participant's right iliac crest, with the reference for the cervical site placed on the left iliac crest. We selected stimulation sites to target cervical and thoracolumbar spinal segments associated with upper- and lower-limb sensorimotor networks [32].

We determined each participant's motor threshold while they remained seated using spinal motor evoked potentials (sMEPs). Motor threshold was defined as the lowest stimulation intensity that elicited a consistent EMG detectable response in at least two recorded proximal lower-limb muscles. Surface electromyography was recorded from the rectus femoris, biceps femoris, and vastus lateralis during threshold testing (2000Hz; Trigno Wireless system, Delsys, Boston, MA, United States). We then set stimulation amplitude for all trials at 50% of each participant's motor threshold [34].

### *Proprioceptive Training with Performance Feedback*

Participants completed proprioceptive training using the Crisscross assessment, which consisted of six training blocks of 18 crossing attempts (108 total attempts; Fig. 1C). A monitor positioned directly in front of the participant displayed test instructions and performance feedback throughout the training session. At the beginning of each crossing attempt, and while the ankles were stationary between crossing attempts, participants were shown the prompt, "Press Button When You Feel Your Feet Are Aligned." Immediately after each button press response, participants received performance feedback displayed as the absolute percent error relative to the total crossing workspace. This feedback informed participants about the magnitude of their error but did not indicate if their response occurred before or after the true crossing position. The prompt then reappeared, initiating the next crossing attempt. If the participant did not respond during a crossing attempt, the prompt remained visible until the next crossing attempt began.

### *Experimental Design*

A cohort of 14 participants completed a single-session protocol to quantify the effects of tSCS on ankle proprioception and gait, as well as the effects of pairing tSCS with proprioceptive training (Fig. 1C). Participants first completed the baseline (BL) proprioception assessment followed by normal (BL NW) and tandem (BL T) walking trials without stimulation. We then applied tSCS, after which participants completed an initial proprioception assessment (PreONa) to quantify the immediate effects of stimulation on proprioception. Participants next completed both normal (PreON-NW) and tandem (PreON-T) walking trials under stimulation, followed by a second proprioception assessment (PreONb) to determine whether walking under stimulation altered proprioception prior to training. Participants then completed six proprioceptive training blocks (T1–T6), followed by a post-training proprioception assessment with stimulation (PostON), post-training normal (PostON-NW) and tandem (PostON-T) walking trials, and a final proprioception assessment after we removed stimulation (PostOFF). Walking trials were used to quantify stimulation- and training-related changes in locomotor control, trunk stability, and whole-body movement dynamics.

Data from a previously collected cohort of 14 participants were included for comparison (Fig. 1C, top row). These participants completed a matched single-session protocol consisting of identical proprioceptive assessments, the same training protocol, and identical performance feedback, but did not undergo gait trials. The control dataset enabled direct comparison of proprioceptive learning trajectories and post-training outcomes between groups. Consequently, changes associated with tSCS can be distinguished from those attributed to repeated practice and performance feedback alone.

### *Proprioceptive Data Processing*

We processed all proprioceptive data in MATLAB (MathWorks, Natick, MA, USA). During the Crisscross assessment, bilateral ankle position trajectories and participant response times were recorded for each crossing attempt. We defined the true crossing event as the time at which both ankles reached the same angular position and determined participant response times from the recorded button-press timestamps.

We quantified proprioceptive performance using three outcome measures: absolute error (AE), timing error (TE), and variable error (VE). AE represented the absolute angular difference between the ankles at the time of the participant's button press.

TE was calculated as the time difference between the participant's button press and the true crossing event. Negative TE values indicated anticipatory responses (button presses occurring before the crossing event), whereas positive values indicated delayed responses (button presses occurring after the crossing event). Because crossing speed was experimentally controlled, TE quantified the temporal accuracy of participants' responses to ankle alignment.

VE was calculated as the standard deviation of the signed error across repeated crossing attempts within each assessment [35]. Because it is computed as the standard deviation about each participant's mean response, VE reflects the consistency of proprioceptive judgments independent of directional bias: a systematic offset in perceived alignment shifts the mean but does not, by itself, increase VE. Lower VE values indicated more consistent judgments across trials, whereas higher values reflected greater trial-to-trial variability.

### *Gait Data Processing*

Gait kinematic data were analyzed in MATLAB. We restricted analyses to the final 60s of each walking trial to capture steady-state walking while excluding gait initiation and termination. Marker trajectories were low-pass filtered at 20Hz to reduce motion artifact (fourth-order, zero-lag Butterworth filter).

We identified gait events using calcaneus marker trajectories relative to the pelvis center, approximated as the average between the anterior superior iliac spine and posterior superior iliac spine markers. We defined heel strike as the local maximum anterior position of the calcaneus marker relative to the pelvis, and toe-off as the local minimum. We segmented continuous gait data into strides using successive right heel strikes and excluded strides with durations outside a self-scaling band around the median stride time.

At heel strike, we calculated step length and step width as the anterior-posterior and mediolateral distance between the calcaneus markers, respectively. We quantified trunk sway from a segment vector extending from the pelvis center to a marker placed on the C7 spinous process. We calculated anterior-posterior (AP) and mediolateral (ML) trunk lean angles as the angle of the trunk vector relative to vertical when projected onto the sagittal and frontal planes, respectively. For each retained stride, we quantified AP and ML trunk sway as the peak-to-peak range of the corresponding trunk lean angle. Mean trunk sway was calculated across all retained strides, with sway variability defined as the stride-to-stride standard deviation of the per-stride range.

We estimated center-of-mass (CoM) motion using the pelvic center trajectory. After segmenting the trajectory into strides, we calculated ML CoM excursion as the peak-to-peak mediolateral displacement of the pelvis center for each stride. We then averaged per-stride excursion values for each trial to obtain a single ML CoM excursion value for each participant and condition.

### Statistics

All statistical analyses were performed using MATLAB (version 2025a, Mathworks) and JMP software (version 17 Pro,SAS Institute Inc., Cary, NC, USA). Statistical significance was defined as $p < 0.05$. For significant main effects, we performed Tukey-adjusted post-hoc comparisons. All linear mixed-effects models included participant as a random effect to account for repeated observations.

*Proprioceptive Analysis*

We fit separate linear mixed-effects models for AE and TE to evaluate the acute effects of tSCS on proprioceptive performance. Fixed effects included stimulation state (ON vs. OFF), crossing speed (16, 26, and 36°/s), and their interaction. Stimulation OFF assessments included BL and PostOFF, whereas simulation ON assessments included PreONa, PreONb, and PostON.

To evaluate proprioceptive learning, we fit separate linear mixed-effects models for AE and TE to determine if proprioceptive performance changed over the training session and if these changes differed between tSCS and control groups. Fixed effects included experimental condition, group (tSCS, control), and their interaction. Because proprioceptive adaptation was not assumed to change linearly across training, experimental condition was treated as a categorical repeated-measures variable. A significant main effect of the experimental condition was interpreted as evidence of learning, whereas a significant experimental condition × group interaction indicated that the pattern of learning differed between groups.

Relationships between proprioceptive performance and dorsiflexion MVC were assessed using Spearman's rank correlation because timing error did not meet the assumption of normality, as indicated by a significant Kolmogorov–Smirnov test ($p < 0.001$).

*Gait Analysis*

We quantified gait performance using step length, step width, trunk sway, and CoM limit cycle area in the anterior-posterior (AP) and mediolateral (ML) directions. We fit separate linear mixed-effects models for each gait outcome measure within each walking task (normal walking and tandem walking). Fixed effects included experimental condition (BL, PreON, and PostON).

## III. Results

### Effects of tSCS on Ankle Proprioception

We evaluated the immediate effects of tSCS on ankle proprioception by comparing Crisscross performance with and without stimulation. For absolute error (AE), the linear mixed-effects model showed a significant main effect of stimulation state ($F = 5.34$, $p = 0.02$). Before training, AE was 9.6 ± 3.9° during BL and 11.1 ± 4.2° during PreONa. After training, AE decreased in both PostON and PostOFF, down to 9.0 ± 2.6° and 7.5 ± 2.8°, respectively. Across assessments, post hoc comparisons showed that stimulation ON (PreON and PostON) increased AE by an average 1.5 ± 2.4° relative to stimulation OFF (BL and PostOFF) ($p < 0.001$, Fig. 2A). There was no effect of stimulation state × speed interaction ($F = 0.18$, $p = 0.80$).

Response timing and response consistency were largely unaffected by stimulation. Timing Error (TE) showed no significant main effect of stimulation state ($F = 2.4$, $p = 0.12$) and no significant effects of crossing speed or stimulation state×speed interaction ($F = 0.7$, $p = 0.5$). Before training, at baseline, average TE was −0.36 ± 0.27 s and −0.30 ± 0.39 s PreONa. After training, at PostOFF TE was −0.06 ± 0.27 s and −0.21 ± 0.26 s at PostON. Thus, stimulation-related changes in signed TE were small and differed across assessment time points, with an average change of −0.05 ± 0.19 s during stimulation ON (PreON and PostON) relative to stimulation OFF (BL and PostOFF; see Fig. 2B). Because positive and negative TE values can partially offset each other, we repeated an additional sensitivity analysis using absolute TE. This analysis revealed a significant main effect of stimulation state ($F = 9.28$, $p = 0.002$), with absolute TE increasing by 0.062 ± 0.11 s during stimulation ON vs stimulation OFF conditions (Fig. 2A). Therefore, although tSCS did not produce a consistent directional shift in response timing, it increased the magnitude of timing errors.

For variable error (VE), no significant main effects or interactions were observed (all $p > 0.30$). Before training, VE was 6.5 ± 2.0° at baseline and 7.4 ± 2.9° for PreON. After training, VE was 7.1 ± 1.9° for PostOFF and 7.7 ± 2.7° during PostON. Across paired assessments, VE increased by only 0.2 ± 2.5° during stimulation ON relative to the corresponding stimulation-OFF assessments, indicating that response variability remained consistent during stimulation (Fig. 2A).

Stimulation-related increases in proprioceptive error were not accompanied by changes in dorsiflexion MVC torque. Dorsiflexion MVC torque was not significantly affected by stimulation (BL: 26.3 ± 5.32 Nm vs. PreONa: 27.4 ± 3.8 Nm; $p = 0.50$). Similarly, stimulation-related changes in AE were not significantly associated with changes in dorsiflexion MVC torque ($\rho = 0.22$, $p = 0.44$; Fig. 2C).

### Effects of Proprioceptive Training

Proprioceptive training produced significant improvements in AE in the tSCS group compared to the control group. The linear mixed-effects model revealed a significant main effect of training block ($F = 5.7$, $p < 0.001$) and a significant group × training block interaction ($F = 2.3$, $p = 0.02$; Fig. 2D).

In the control group, AE remained relatively unchanged across the training session, with no significant reductions from BL at any training block or during the post-proprioceptive training assessments (all p > 0.20; Fig. 2D). In contrast, participants who trained with tSCS demonstrated progressive improvements in proprioceptive accuracy during the later stages of training. AE decreased by 1.92 ± 3.5° at training block 6 (T6) relative to BL (p = 0.002) and remained 2.1° ± 4.2 lower than BL during the post-training PostOFF assessment (p = 0.003; Fig. 2D). PostOFF performance did not differ from training block 6 (p = 0.8), indicating that these improvements persisted even after stimulation was removed. Although AE did not differ significantly between the tSCS and control groups at any individual training block or at the PostOFF assessment (all p > 0.05), the significant group × training block interaction indicated that the pattern of change across training differed between groups. Specifically, AE decreased from baseline in the tSCS group but remained relatively unchanged in the control group.

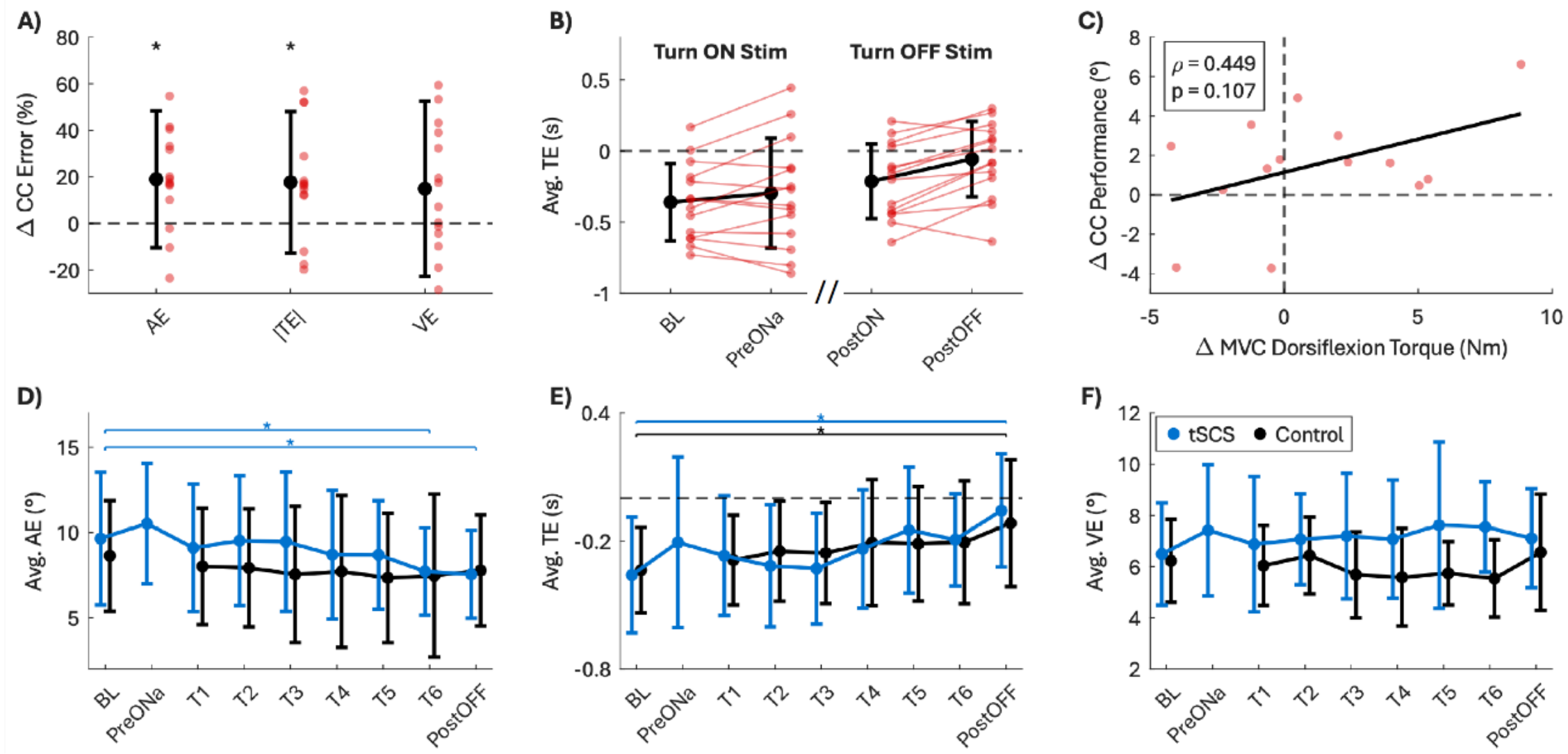

**Figure 2.** Effects of tSCS on conscious ankle proprioception and its change with training. **A.** Mean and individual participant changes in proprioceptive error during tSCS, expressed as the percent difference between stimulation-ON and stimulation-OFF assessments for absolute error (AE), variable error (VE), and absolute timing error (|TE|). **B.** Average timing error (TE) across paired stimulation assessments in the tSCS group, showing changes when stimulation was turned on and turned off. **C.** Relationship between stimulation-induced changes in Crisscross proprioceptive performance and changes in dominant-limb (Dom) dorsiflexion MVC torque. **(D–F)** Average performance in AE (D), TE (E), and VE (F) across the training protocol for the tSCS and control group. The PreONa assessment is plotted to display the raw baseline data but was not included in the linear mixed-effects models; significance bars therefore denote comparisons among the modeled training assessments only. *p < 0.05.

Response timing changed significantly over the training period. The linear mixed-effects model revealed a significant main effect of experimental condition (F = 26.3, p < 0.001) and a significant experimental condition × group interaction (F = 2.7, p = 0.01). Across the protocol, TE shifted toward less anticipatory responses, with post-training performance shifting closer to zero compared with baseline (BL: −0.25 ± 0.24 s; PostOFF: −0.14 ± 0.30 s; p < 0.001; Fig. 2E). However, no significant between-group comparisons were observed at any individual experimental condition (all p > 0.50), indicating that both groups demonstrated similar changes in timing throughout training.

Changes in response variability (VE) were not significantly affected by training or tSCS. The linear mixed-effects model revealed no significant main effects of group (F = 0.11, p = 0.736) or experimental condition (F = 0.40, p = 0.902), and no significant group × experimental condition interaction (F = 1.21, p = 0.302; Fig. 2F).

### *Acute and Training Effects of tSCS on Normal Walking*

Acute tSCS led to a more constrained walking pattern, with reductions in spatiotemporal, trunk, and CoM-based measures (Fig. 3 Top). Normalized step length decreased from 54.7 ± 2.5 at baseline (BL-NW) to 54.4 ± 2.6 (dimensionless) during PreON-NW, corresponding to a 0.5 ± 2.0% reduction (p < 0.001). Normalized step width decreased from 13.9 ± 3.4 at baseline to 12.4 ± 3.6 (dimensionless) during PreON-NW, (11.6 ± 8.8%; p < 0.001). AP trunk sway decreased by 3.5 ± 6.5% (BL-NW: 3.8 ± 0.5°; PreON-NW: 3.7 ± 0.5°; *p* < 0.001) and ML trunk sway decreased by 7.6 ± 16.9% (BL-NW: 2.5 ± 1.1°; PreON-NW: 2.3 ± 1.0°; p < 0.001). Similarly, ML CoM excursion decreased by 6.6 ± 6.4% relative to baseline (BL-NW: 61.9 ± 12.6 mm; PreON-NW: 57.8 ± 12.1 mm; p < 0.001).

Following proprioceptive training with tSCS, several sagittal-plane gait measures recovered toward or exceeded baseline values. Step length increased from 54.4 ± 2.6 to 54.9 ± 2.6 (dimensionless) from PreON-NW to PostON-NW, (0.88 ± 1.0%; p < 0.001), also exceeding baseline values (p = 0.005). AP trunk sway, similarly, returned to comparable baseline values (p > 0.20). Thus, the acute reductions in sagittal-plane gait dynamics largely resolved following proprioceptive training.

In contrast, frontal-plane gait measures remained below baseline after training. Step width remained significantly lower than baseline during PostON-NW (11.6 ± 2.8 (dimensionless); p < 0.001) and was further reduced relative to PreON-NW (p < 0.001). ML trunk sway increased following training (PostON-

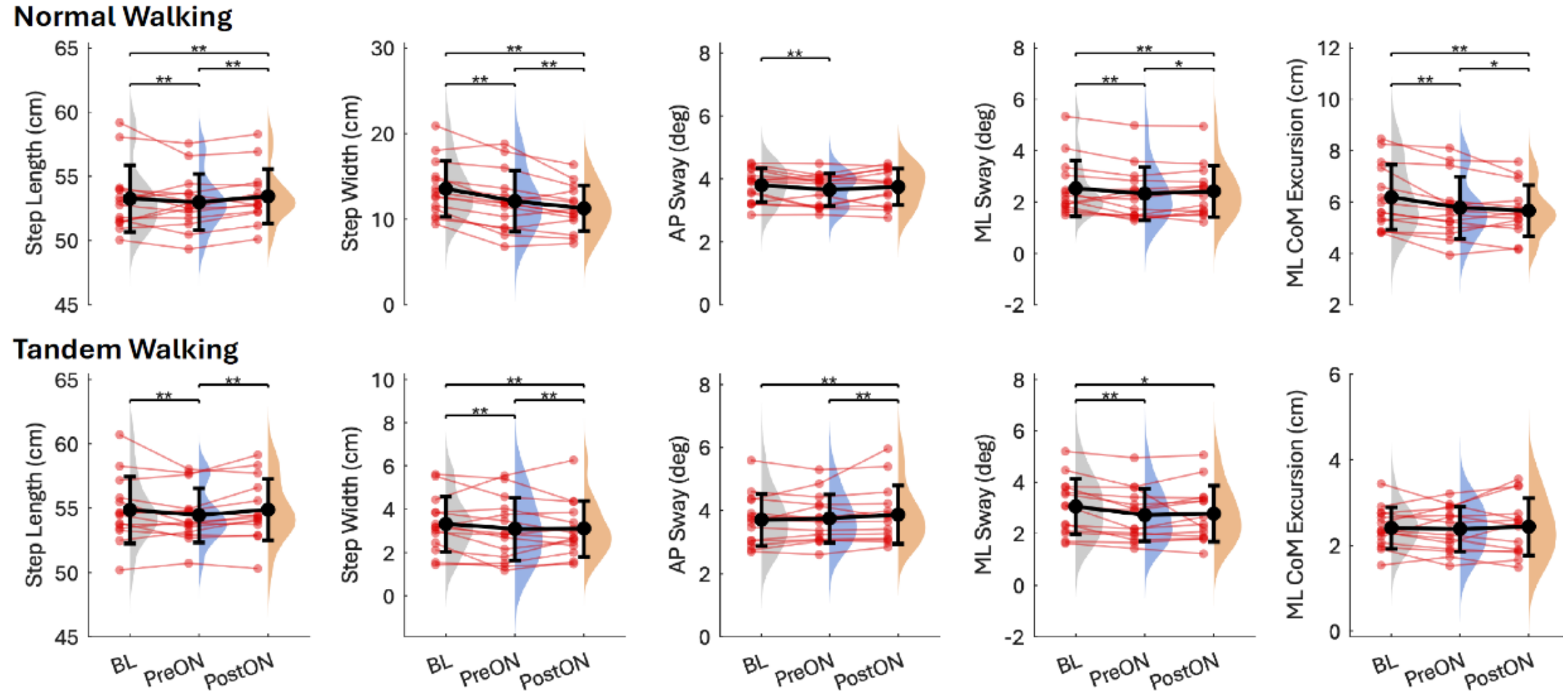

**Figure 3.** Effects of tSCS and proprioceptive training on gait during normal and tandem walking. **Top Row:** Gait outcomes for normal walking across baseline (BL), pre-training stimulation ON (PreON), and post-ankle proprioception training with stimulation ON (PostON). **Bottom Row:** Gait outcomes for tandem walking across BL, PreON, and PostON. Red points and lines represent individual participants, black points and error bars represent group mean ± SD, and shaded distributions show the spread of participant values at each time point.

NW: 2.4 ± 1.0°; p = 0.01 when compared to PreON-NW) but still remained significantly below baseline ($p < 0.01$). ML CoM excursion also moved toward baseline following training (PostON-NW: 56.6±10.0mm, p = 0.01) but was still significantly below baseline ($p < 0.01$). These findings indicate that training, or time of exposure to tSCS, restored sagittal-plane gait dynamics, whereas frontal-plane gait remained constrained during stimulation.

Stride-to-stride variability was less affected by stimulation and training than mean gait behavior. Although the mixed-effects models identified a significant effect of experimental condition on several measures of variability, none of the Tukey-adjusted pairwise comparisons were significant (all $p > 0.05$). Descriptively, variability in step width, step length, and trunk sway showed trends consistent with increases during stimulation and recovery toward baseline following proprioceptive training. ML CoM excursion variability remained unchanged throughout the protocol.

### *Acute and Training Effects of tSCS on Tandem Walking*

Acute tSCS application resulted in a more constrained walking pattern during tandem walking compared to baseline, similar to the effect observed during normal walking (Fig. 3, bottom). Normalized step length decreased by 0.7 ± 1.7% relative to baseline (BL-T: 56.3 ± 2.9; PreON-T: 55.9 2.8 (dimensionless); $p < 0.001$), while normalized step width decreased by 8.2 ± 19.2% (BL-T: 3.4 ± 1.4; PreON-T: 3.4 ± 1.5 (dimensionless); $p < 0.001$). ML trunk sway also decreased by 10.6 ± 9.3% (BL-T: 3.1± 1.1°; PreON-T: 2.7± 1.0°; $p < 0.001$). AP trunk showed a non-significant trend consistent with an increase after stimulation (BL-T: 3.7 ± 0.8°; PreON-T: 3.7 ± 0.7; p = 0.4). ML CoM excursion remained unchanged across experimental conditions (F = 0.91, p = 0.4).

Following proprioceptive training with tSCS, several sagittal-plane gait measures recovered toward or exceeded baseline values. Step length increased by 0.8 ± 1.2% from PreON-T to PostON-T (55.9 ± 2.8 and 56.3 ± 2.8 (dimensionless), respectively; $p < 0.001$), with PostON-T values not differing from BL-T (p = 0.8). AP trunk sway also increased following training (PostON-T: 3.9 ± 0.9°), increasing by 3.0 ± 6.9% relative to PreON-T (p = 0.001) and exceeding BL-T values by 4.6 ± 10.5% ($p < 0.001$). These findings suggest that, similar to normal walking, training restored sagittal-plane dynamics during tandem walking.

In contrast, frontal-plane gait measures remained altered following training. Step width increased by 6.2 ± 20.9% from PreON-T to PostON-T (3.2± 1.5 and 3.2 ± 1.4 (dimensionless), respectively; $p < 0.001$), but remained 3.4 ± 25.9% below BL-T ($p < 0.001$). ML trunk sway did not differ between PreON-T: and PostON-T (2.7± 1.0° and 2.8 ± 1.1°, respectively; p = 0.5), but remained significantly below BL-T ($p < 0.05$). Similarly, ML CoM excursion did not change across experimental conditions (F = 0.91, p = 0.4).

Stride-to-stride variability measures were largely unaffected by stimulation or training. No significant differences were observed in step-length variability, step-width variability, AP trunk-sway variability, or ML trunk-sway variability across experimental conditions (all $p > 0.10$).

## IV. Discussion

The purpose of this study was to determine how modulation of afferent input through tSCS influences conscious ankle proprioception, locomotor behavior, and sensorimotor adaptation in unimpaired adults. Consistent with our hypothesis, acute tSCS impaired conscious ankle proprioceptive performance, demonstrating that increasing afferent input does not necessarily improve sensory accuracy. In response, participants adopted shorter, narrower steps with reduced trunk motion and ML CoM excursions during both normal and tandem walking. Following repeated exposure to tSCS during proprioceptive training, ankle proprioceptive accuracy returned toward baseline values, and several gait measures returned near or beyond baseline values. Thus, the nervous system adapted to the altered sensory environment induced by tSCS. Collectively, the results demonstrate that spinal neuromodulation differentially influences multiple aspects of the sensorimotor control loop, affecting conscious proprioceptive perception, locomotor behavior, and sensorimotor adaptation.

### tSCS Disrupts Ankle Proprioception

As we hypothesized, acute tSCS impaired ankle proprioceptive accuracy during the Crisscross assessment. This may appear counterintuitive given that tSCS is widely reported to enhance motor output and locomotor performance [11], [17], [18], [19], [20]. However, these results indicate that enhanced afferent recruitment does not necessarily translate into more accurate sensory perception.

Importantly, the reduction in proprioceptive accuracy occurred without changes in gross motor output, as dorsiflexion MVC torque remained unchanged during stimulation. Moreover, tSCS did not affect response consistency or variable error. Similarly, timing error showed only a modest, non-significant shift toward less anticipatory responses. The magnitude of timing error (|TE|), however, increased in parallel with AE. Because signed timing errors partially cancel across anticipatory and reactive responses, mean timing error remained near zero even as |TE| grew [25]. This suggests that tSCS increased the imprecision of response timing, because Crisscross converts timing error into spatial error, manifested as larger errors in perceived ankle position.

One mechanism that may contribute to the reduction in proprioceptive accuracy is alteration of the proprioceptive information provided to the spinal and supraspinal sensorimotor networks. tSCS primarily recruits large-diameter dorsal root afferents and modulates spinal sensorimotor networks through activation of sensory pathways [32], [36], [37], [38]. Although increased afferent activity may facilitate motor function by increasing the excitability of sensorimotor circuits, it may simultaneously alter how proprioceptive signals are temporally encoded, weighted, and integrated by the nervous system. Consequently, the sensory information available for estimating limb position and movement may no longer align with established sensorimotor predictions and internal representations of limb position, reducing the accuracy of conscious proprioceptive perception without creating motor impairment, such as reduced strength.

A second potential mechanism that may lead to changes in proprioceptive accuracy involves antidromic collisions within stimulated afferent pathways. Activation of large-diameter dorsal root afferents by spinal stimulation generates both orthodromic and antidromic action potentials [24]. Antidromic activity propagating toward the periphery may interfere with normally occurring proprioceptive signals generated during ankle movement, preventing some afferent activity from reaching the spinal cord and reducing the fidelity of proprioceptive signaling while leaving gross motor output relatively maintained. This mechanism is supported by Formento et al. who demonstrated that epidural spinal stimulation can disrupt proprioceptive transmission through antidromic collisions, reducing conscious perception of leg position and impairing detection of passive leg movement [24]. They concluded that preserving proprioceptive information is critical for effective locomotor control. The transcutaneous stimulation used in the present study may recruit a broader population of afferent pathways than epidural stimulation. This may therefore increase the potential for interference with normally occurring sensory signals. Although antidromic collisions were not directly assessed in the present study, the observed reduction in proprioceptive accuracy is consistent with the proposed effects of antidromic activity on proprioceptive signaling.

A third potential mechanism involves increased neural noise within proprioceptive pathways. The Crisscross test requires continuous estimation of the relative positions of both ankles and the precise moment when the limbs are aligned. Accurate performance, therefore, is highly dependent on reliable proprioceptive state estimation [3]. Artificial activation of sensory pathways may increase variability in afferent activity, thereby increasing uncertainty in the proprioceptive information used to estimate limb position and movement [39]. Greater sensory uncertainty could reduce the accuracy of state estimation and proprioceptive judgments, even when motor output remains intact. This interpretation is consistent with the observed increase in ankle proprioceptive absolute error despite no change in dorsiflexion MVC.

However, variable error remained unchanged throughout stimulation, indicating that tSCS did not increase variability in proprioceptive performance. Instead, participants made consistently larger errors in ankle position estimation. Consistent with this, absolute timing error increased in parallel, reflecting the tight coupling between timing and spatial error in the Crisscross assessment [25]. This distinction suggests that tSCS reduced the accuracy, but not the consistency, of conscious proprioceptive perception. Previous studies have similarly shown that peripheral sensory stimulation, such as tendon vibration or subsensory electrical stimulation, can alter proprioceptive perception by modifying the signal-to-noise characteristics of afferent input, thereby influencing the sensory estimates used for perception and motor control [39], [40], [41].

### tSCS Leads to a More Constrained Gait Pattern

A major finding of the present study was that multiple gait characteristics, step length, step width, trunk sway, and ML CoM excursion, changed when conscious ankle proprioception was disrupted, even though participants remained able to complete both normal and tandem walking. Although many of these changes reached statistical significance, several were small in absolute magnitude (e.g., step-length changes $< 1\%$) and were detected within a repeated-measures design with many strides per participant, which affords high sensitivity to small mean shifts. The strength of this finding, therefore, rests less on the magnitude of any single measure than on the consistent convergence of step width, trunk sway, and ML CoM excursion toward reduced values, with the largest and most reliable effects appearing in frontal-plane measures during normal walking. Importantly, this convergence occurred without any corresponding increase in gait variability: step length and step width variability were unchanged, and trunk sway variability showed only small, non-significant reductions.

The coordinated reductions in trunk sway, ML CoM excursion, and step width suggest that tSCS shifted locomotion toward a lower-excursion, narrower-base pattern rather than producing generalized gait destabilization. This coupling, in which reduced trunk and CoM motion is accompanied by a narrower base of support, has been observed when healthy adults are instructed to restrict trunk motion during walking, where the narrower base preserves a constant mediolateral margin of stability [42], and in the narrow-based gait associated with reduced trunk motion in Parkinson's disease [43]. Notably, this pattern is the opposite of that produced by frontal-plane destabilization, which increases both CoM sway and step width [44]. The present findings are therefore more consistent with a coordinated downscaling of locomotor excursion, particularly in the frontal plane, than with a cautious or destabilized gait, which is typically characterized by wider steps and greater variability [45], [46]. Because a narrower base does not by itself indicate greater stability, direct measures such as the margin of stability, local dynamic stability, or responses to perturbation would be required to determine the functional consequences of this pattern.

The modest magnitude of the observed kinematic changes is consistent with locomotor control drawing on multiple sensory and motor pathways that can compensate for temporary disruptions in conscious proprioceptive perception [47], [48], [49]. Whereas the Crisscross test requires explicit awareness of ankle position and movement, walking depends on integrating visual [50], vestibular [50], cutaneous [49], and proprioceptive information [51], [52] across spinal and supraspinal sensorimotor networks. Therefore, a reduction in the accuracy of conscious proprioceptive judgments may not necessarily impair the generation of coordinated locomotor output. Previous work has similarly shown that locomotion can remain remarkably robust despite alterations in sensory feedback, particularly when other sensory channels remain available to support movement control [53], [54], [55].

This redundancy is formalized in the concept of sensory reweighting: when the reliability of one sensory source is altered, the nervous system can modify the relative weighting of available visual, vestibular, and somatosensory information to maintain locomotor performance [48], [56]. Under this framework, stimulation-induced alterations in proprioceptive signaling may have reduced the weighting of ankle proprioceptive information while increasing reliance on other sensory inputs or predictive motor mechanisms. Such compensatory processes may explain why gait was minimally affected despite measurable changes in proprioceptive accuracy.

Interestingly, the effects of stimulation were not uniform across gait outcomes. During normal walking, reductions in both AP and ML locomotor dynamics were observed immediately following stimulation. Following proprioceptive training, however, AP measures such as step length and AP trunk sway recovered toward baseline values, whereas reductions in step width, ML sway, and ML CoM excursion persisted. A similar pattern emerged during tandem walking, where AP locomotor measures recovered and, in some cases, exceeded baseline values, while frontal-plane measures remained relatively constrained. These findings suggest that the sensorimotor system adapted differently across locomotor domains, with sagittal-plane progression recovering more readily than frontal-plane control.

This directional dissociation aligns with an established principle of bipedal balance: the sagittal and frontal planes are stabilized by different mechanisms with different sensory demands. Fore–aft motion during walking is largely passively stable or maintained by low-level spinal feedback, whereas lateral motion is dynamically unstable and requires active, supraspinally mediated feedback control, implemented primarily through regulation of lateral foot placement [57], [58]. Because lateral balance depends more heavily on continuous integration of sensory feedback, it could be disproportionately sensitive to a stimulation-induced degradation of proprioceptive signaling. Consistent with this, AP measures recovered after training, while ML measures (step width, ML sway, ML CoM excursion) remained constrained, suggesting that sagittal-plane control, able to rely on passive dynamics and spinal feedback, readily renormalized, whereas frontal-plane control, more dependent on the altered afferent channel, remained reorganized. The fixed treadmill speed (1 m/s) may further accentuate this asymmetry by externally constraining fore–aft progression, leaving the mediolateral domain as the primary axis in which an altered control strategy could persist.

Collectively, these results show that tSCS induced a reorganization of gait characterized by altered movement patterns and reduced locomotor excursion, while proprioceptive accuracy was concurrently reduced.

### *Proprioceptive Training Facilitated Proprioceptive Improvements During Neuromodulation*

Although the acute application of tSCS impaired ankle proprioceptive accuracy, repeated exposure to the stimulation during proprioceptive training resulted in progressive improvements in performance. Following training, proprioceptive error returned toward baseline values despite continued stimulation, and performance after stimulation removal was comparable to baseline values. These findings indicate that the nervous system adapted to the altered sensory environment created by tSCS; thus, the initial proprioceptive disruption was not permanent or irreversible.

One possible explanation is that repeated exposure to altered sensory input promoted sensorimotor recalibration. The central nervous system continuously updates estimates of body position and movement by combining predicted sensory consequences of movement with incoming sensory feedback, and discrepancies between these signals can drive sensorimotor adaptation and proprioceptive recalibration [59], [60], [61]. During the early stages of stimulation, altered afferent signaling may have produced a mismatch between incoming proprioceptive information and established sensorimotor predictions, resulting in increased error. However, with repeated task practice and performance feedback, participants may have gradually recalibrated these sensory estimates, reducing the discrepancy between perceived and actual ankle position. Similar processes have been observed during visuomotor adaptation [62], force-field adaptation [63], and proprioceptive perceptual learning paradigms [25], [64], [65], where repeated exposure to altered sensorimotor mappings or mechanical environments leads to progressive reductions in error and improvements in performance, with time courses similar to what we observed here.

The inclusion of performance feedback during the Crisscross test likely contributed to this adaptation process. Following each crossing attempt, participants received information regarding the magnitude of their error, allowing sensory estimates to be continuously updated throughout training. Feedback-driven improvements in proprioceptive acuity have been reported previously in both healthy and neurologically impaired populations and are thought to involve plastic changes across spinal, cerebellar, and cortical sensorimotor networks [5], [66], [67]. The present findings suggest that these adaptive mechanisms remain effective even when sensory processing is altered through spinal neuromodulation.

Interestingly, improvements in proprioceptive accuracy occurred without a corresponding reduction in response variability. This pattern suggests that training primarily recalibrated the average proprioceptive estimate rather than reducing trial-to-trial uncertainty. This distinction is important because variability within sensorimotor systems is not simply noise or error; it can reflect ongoing exploration, flexibility, and sampling of sensory states during learning [68], [69]. In conjunction with proprioceptive changes, gait measures also changed over the course of the session. During normal walking, sagittal-plane measures such as step length and AP trunk sway returned toward baseline following training. A similar pattern occurred during tandem walking, where step length and AP trunk sway increased and, in some cases, exceeded baseline. In contrast, reductions in ML trunk sway and ML CoM excursion persisted.

These findings highlight the importance of considering adaptation and learning when evaluating neuromodulation-based interventions, as short-term responses may differ substantially from the changes that emerge following repeated practice over a longer period.

### *Implications for Neuromodulation and Sensorimotor Control*

Although the current study focused on individuals without gait impairments, the findings have implications for the interpretation and application of spinal neuromodulation in rehabilitation. Most studies investigating tSCS have focused primarily on motor outcomes, including muscle activation, standing ability, and locomotor performance [11], [17], [18], [19], [20]. Improvements in these measures are often interpreted as evidence that stimulation enhances sensorimotor function through increased afferent recruitment and spinal network excitability. However, the current results demonstrate that tSCS can also disrupt conscious proprioceptive accuracy, resulting in immediate changes to gait.

The distinction between acute and training-related effects may be particularly important. Immediately following the onset of stimulation, proprioceptive accuracy declined, while locomotor behavior shifted toward a more constrained movement pattern. Following repeated exposure to the altered sensory environment accompanied by error feedback, proprioceptive performance improved, and several locomotor measures recovered toward or exceeded baseline values. These observations indicate that the immediate negative effects of neuromodulation on conscious proprioceptive accuracy may be compensated through training. Therefore, acute responses to stimulation may not reliably predict longer-term therapeutic outcomes, as substantial sensorimotor adaptation can emerge with repeated exposure and training.

The present results also highlight the importance of considering sensory processing as a therapeutic target rather than solely focusing on motor performance. Proprioceptive impairments are common following neurological injury and are associated with deficits in balance, gait, and functional mobility [70], [71], [72]. While spinal stimulation is frequently viewed as a tool for enhancing motor output, the current findings suggest that it may also alter how sensory information is processed and integrated. Understanding these sensory effects may be particularly important when designing interventions intended to improve locomotor recovery, as changes in afferent processing could influence both movement execution and motor learning.

More broadly, the findings support the concept that spinal neuromodulation influences multiple aspects of the sensorimotor control loop [14], [32], [73]. This perspective may help explain why individuals often exhibit variable responses to neuromodulation and why improvements in one domain do not always correspond to improvements in another. Future rehabilitation strategies may therefore benefit from incorporating both sensory and motor outcome measures to better characterize the mechanisms through which neuromodulation promotes recovery.

### *Limitations and Future Directions*

Several limitations should be noted. First, the sample was small and warrants replication in a larger cohort. Second, proprioception was evaluated with a single assessment (Crisscross), which measures conscious perception of passive dynamic ankle alignment. Proprioception is multifaceted; therefore, other assessments such as joint position reproduction, movement detection thresholds, force sense, or weight-bearing assessments may respond differently. Third, gait trials were not performed in the control cohort, so we could not separate the locomotor adaptations after training from stimulation, proprioceptive training, or nonspecific practice effects. Fourth, the study was conducted in young unimpaired adults, and thus the extent to which the findings generalize to neurological populations remains to be determined. Individuals with stroke or spinal cord injury often exhibit baseline proprioceptive deficits, altered locomotor control, and varied responses to neuromodulation, therefore the effects of SCS on conscious proprioception remain of high interest. Finally, we examined one set of stimulation parameters. Because neuromodulation effects depend on stimulation location [32], intensity [74], and frequency [75], future work should test how these sensory and locomotor adaptations vary across stimulation configurations.

## V. Conclusion

Acute tSCS impaired conscious ankle proprioceptive perception and immediately altered locomotor kinematics. Gait parameters shifted toward a modest but consistent reduction in locomotor excursion, characterized by reduced step width, trunk sway, and CoM limit cycle area, without an increase in gait variability. Following proprioceptive training with stimulation applied, proprioceptive error returned toward control values and gait recovered in a direction-dependent manner, with sagittal-plane measures renormalizing while mediolateral measures remained constrained. These results underscore the potential importance of accounting for proprioceptive effects in stimulation-based rehabilitation and motivate work on the underlying mechanisms and their impact on recovery in neurologically impaired populations.

## Acknowledgment

This paper is dedicated to the memory of our colleague and mentor Dr. V. Reggie Edgerton, who contributed to this work but passed away before its completion. His insight and his foundational contributions to the field profoundly shaped this study, and he is deeply missed.